\documentclass[fleqn]{article}

\usepackage{amsmath}
\usepackage{amssymb}
\usepackage{cite}
\usepackage{hyperref}

\begin{document}

\title{Gauge-invariant description of some (2+1)-dimensional \\ integrable nonlinear evolution equations}

\author{V.~G. Dubrovsky and A.~V. Gramolin
\\{\small Novosibirsk State Technical University, Karl Marx prosp. 20, Novosibirsk 630092, Russia}
\\{\small E-mail: \href{mailto:dubrovsky@academ.org}{dubrovsky@academ.org} and \href{mailto:gramolin@gmail.com}{gramolin@gmail.com}}}

\date{}

\maketitle

\begin{abstract}
New manifestly gauge-invariant forms of two-dimensional generalized dispersive long-wave and Nizhnik--Veselov--Novikov systems of integrable nonlinear equations are presented. It is shown how in different gauges from such forms famous two-dimensional generalization of dispersive long-wave system of equations, Nizhnik--Veselov--Novikov and modified Nizhnik--Veselov--Novikov equations and other known and new integrable nonlinear equations arise. Miura-type transformations between nonlinear equations in different gauges are considered. 
\end{abstract}

\section{Introduction}
\label{Section_1}

The fundamental ideas of gauge invariance and gauge transformations are wide spread and in common use in almost every part of physics. The first applications of such ideas in the theory of integrable nonlinear equations by Zakharov and Shabat~\cite{Zakharov&Shabat_1974}, Kuznetsov and Mikhailov~\cite{Kuznetsov&Mikhailov_1977}, Zakharov and Mikhailov~\cite{Zakharov&Mikhailov_1978}, Zakharov and Takhtadzhyan~\cite{Zakharov&Takhtajan_1979}, Konopelchenko~\cite{Konopelchenko_PLA_1982}, Konopelchenko and Dubrovsky~\cite{Konopelchenko&Dubrovsky_1983, Konopelchenko&Dubrovsky_1984} and others have been made, see also the books~\cite{Theory_of_solitons, Faddeev&Takhtajan, Ablowitz&Clarkson, Konopelchenko_book_1, Konopelchenko_book_2, Konopelchenko_book_3} and references therein.

Now a lot of gauge-equivalent to each other, integrable nonlinear models are well known. In one-dimensional case the most famous are the nonlinear Schr\"odinger and Heisenberg ferromagnet equations, massive Thirring model and two-dimensional relativistic field model, KdV and mKdV equations and so on; in the two-dimensional case the most famous are Kadomtsev--Petviashvili and modified Kadomtsev--Petviashvili nonlinear equations, Davey--Stewartson and Ishimori integrable systems of nonlinear equations and so on. See some references in the books~\cite{Theory_of_solitons,Faddeev&Takhtajan, Ablowitz&Clarkson, Konopelchenko_book_1, Konopelchenko_book_2, Konopelchenko_book_3, Konopelchenko&Rogers_1992}.

In the present paper, manifestly gauge-invariant formulation of two-dimensional nonlinear evolution equations integrable by the following two scalar auxiliary linear problems:
\begin{gather}
L_1 \psi = \bigl(\partial_{\xi \eta}^2 + u_1 \partial_{\xi} + v_1 \partial_{\eta} + u_0\bigr) \psi = 0, \label{first_auxiliary} \\
L_2 \psi = \bigl(\partial_t + u_3 \partial_{\xi}^3 + v_3 \partial_{\eta}^3 + u_2 \partial_{\xi}^2 + v_2 \partial_{\eta}^2 + \tilde u_1 \partial_{\xi} + \tilde v_1 \partial_{\eta} + v_0\bigr) \psi = 0 \label{second_auxiliary}
\end{gather}
is developed. Here as usual $\xi = x + \sigma y$, $\eta = x - \sigma y$, $\sigma^2 = \pm 1$ and $\partial_{\xi} = \partial / \partial \xi$, $\partial_{\eta} = \partial / \partial \eta$, $\partial_{\xi}^2 = \partial^2 / \partial \xi^2$, etc.

Two cases of auxiliary linear problems (\ref{first_auxiliary}), (\ref{second_auxiliary}) with different second auxiliary linear problem (\ref{second_auxiliary}) are studied:
\begin{itemize}
\item (i) $u_3 = \kappa_1 = \mbox{const}$, $v_3 = \kappa_2 = \mbox{const}$, third-order problem $L_2\psi=0$, such choice of second auxiliary problem (\ref{second_auxiliary}) leads to famous Nizhnik--Veselov--Novikov (NVN)~\cite{Nizhnik_1980, Veselov&Novikov_1984}, modified Nizhnik--Veselov--Novikov (mNVN) \cite{Konopelchenko_RMP_1990} and other equations;
\item (ii) $u_3 = v_3 = 0$, $u_2 = \kappa_1 = \mbox{const}$, $v_2 = \kappa_2 = \mbox{const}$, second-order problem $L_2\psi=0$, such choice of second auxiliary problem (\ref{second_auxiliary}) leads to famous two-dimensional generalization of dispersive long-wave equation (2DDLW)~\cite{Boiti&Leon&Pempinelli_1987}, Davey--Stewartson (DS) system of equations \cite{Davey&Stewartson_1974} and its reductions and other equations.
\end{itemize}

All above-mentioned famous integrable nonlinear equations via the compatibility condition of auxiliary linear problems (\ref{first_auxiliary}) and (\ref{second_auxiliary}) in the form of Manakov's triad representation~\cite{Manakov_UMN_1976}
\begin{equation}
[L_1, L_2] = B L_1 \label{introd_compatibility_condition}
\end{equation}
have been previously established~\cite{Nizhnik_1980, Veselov&Novikov_1984, Konopelchenko_RMP_1990, Boiti&Leon&Pempinelli_1987}, see also books~\cite{ Konopelchenko_book_2, Konopelchenko_book_3} and references therein.

In the paper, gauge transformations
\begin{equation}
\psi \rightarrow \psi' = g^{-1} \psi \label{gauge_transform_introd}
\end{equation}
with arbitrary gauge function $g(\xi, \eta, t)$ of auxiliary linear problems (\ref{first_auxiliary}) and (\ref{second_auxiliary}) are studied. The convenient for gauge-invariant formulation field variables, classical gauge invariants $w_2$, $\widetilde w_2$, $w_1$,
\begin{gather}
w_2 \stackrel{\mathrm{def}}{=} u_0 - u_{1\xi} - u_1 v_1 = u_0' - u_{1\xi}' - u_1' v_1', \label{w2_introd} \\
\widetilde w_2 \stackrel{\mathrm{def}}{=} u_0 - v_{1\eta} - u_1 v_1 = u_0' - v_{1\eta}' - u_1' v_1', \label{tildew2_introd} \\
w_1 \stackrel{\mathrm{def}}{=} u_{1\xi} - v_{1\eta} = u_{1\xi}' - v_{1\eta}' \label{w1_introd}
\end{gather}
and pure gauge variable $\rho$ connected with field variable $u_1(\xi,\eta,t)$ by the formula
\begin{equation}
u_1 \stackrel{\mathrm{def}}{=} \left(\ln{\rho}\right)_{\eta} \label{ro_introd}
\end{equation}
are introduced. The variable $\rho$ corresponds to pure gauge degrees of freedom and has under~(\ref{gauge_transform_introd}) the following simple law of transformation:
\begin{equation}
\rho \rightarrow \rho' = g \rho. \label{rhoTrLaw_Introd}
\end{equation}

Let us mention that for the first auxiliary linear problem (\ref{first_auxiliary}), considered as classical partial differential equation, the invariants $w_2$ and $\widetilde w_2$ from the early times (see for example the classical book of Forsyth~\cite{Forsyth_book}) as Laplace invariants $h = w_2$ and $k = \widetilde w_2$ are known.

The main results of the paper are the following new integrable systems of nonlinear equations in terms of field variables $\rho, w_1, w_2$ given by (\ref{w2_introd})--(\ref{ro_introd}).

In the case (i) of third-order linear auxiliary problem (\ref{second_auxiliary}) the first invariant $w_1$ is equal to zero $w_1 \equiv 0$ and the established integrable system of nonlinear equations in terms of $\rho$, $w_2$ has the form
\begin{gather}
\rho_t = -\kappa_1 \rho_{\xi \xi \xi} - \kappa_2 \rho_{\eta \eta \eta} - 3\kappa_1 \rho_{\xi} \partial_{\eta}^{-1} w_{2\xi} - 3\kappa_2 \rho_{\eta} \partial_{\xi}^{-1} w_{2\eta} + v_0 \rho, \label{rho_t_i_introd} \\
w_{2t} = -\kappa_1 w_{2\xi \xi \xi} - \kappa_2 w_{2\eta \eta \eta} - 3\kappa_1 \bigl(w_2 \partial_{\eta}^{-1} w_{2\xi}\bigr)_{\xi} - 3\kappa_2 \bigl(w_2 \partial_{\xi}^{-1} w_{2\eta}\bigr)_{\eta}. \label{w2t_i_introd}
\end{gather}
It is remarkable that the gauge-invariant subsystem of the system (\ref{rho_t_i_introd})--(\ref{w2t_i_introd}), equation (\ref{w2t_i_introd}) for the gauge invariant $w_2 = u_0 - u_{1\xi} - u_1 v_1$, coincides in form with the famous NVN equation \cite{Nizhnik_1980, Veselov&Novikov_1984}
\begin{equation}
u_{t} = -\kappa_1 u_{\xi \xi \xi} - \kappa_2 u_{\eta \eta \eta} - 3\kappa_1 \bigl(u \partial_{\eta}^{-1} u_{\xi}\bigr)_{\xi} - 3\kappa_2 \bigl(u \partial_{\xi}^{-1} u_{\eta}\bigr)_{\eta}.
\end{equation}

Due to the last remark the system (\ref{rho_t_i_introd})--(\ref{w2t_i_introd}) will be named below as the Nizhnik--Veselov--Novikov (NVN) system of equations.

In the case (ii) of second-order linear auxiliary problem (\ref{second_auxiliary}) the established integrable system of nonlinear equations in terms of $\rho$, $w_1$ and $w_2$ has the form
\begin{gather}
\rho_t = -\kappa_1 \rho_{\xi \xi} - \kappa_2 \rho_{\eta \eta} - 2\kappa_1 \rho \partial_{\eta}^{-1} w_{2\xi} + 2\kappa_2 \rho_{\eta} \partial_{\xi}^{-1} w_1 + v_0 \rho, \label{rho_t_ii_introd} \\
w_{1t} = -\kappa_1 w_{1\xi \xi} + \kappa_2 w_{1\eta \eta} - 2\kappa_1 w_{2\xi \xi} + 2\kappa_2 w_{2\eta \eta} - 2\kappa_1 \bigl(w_1 \partial_{\eta}^{-1} w_1\bigr)_{\xi} + 2\kappa_2 \bigl(w_1 \partial_{\xi}^{-1} w_1\bigr)_{\eta}, \label{w1t_ii_introd} \\
w_{2t} = \kappa_1 w_{2\xi \xi} - \kappa_2 w_{2\eta \eta} - 2\kappa_1 \bigl(w_2 \partial_{\eta}^{-1} w_1\bigr)_{\xi} + 2\kappa_2 \bigl(w_2 \partial_{\xi}^{-1} w_1\bigr)_{\eta}. \label{w2t_ii_introd}
\end{gather}

The gauge-invariant subsystem of the system (\ref{rho_t_ii_introd})--(\ref{w2t_ii_introd}), the system of equations~(\ref{w1t_ii_introd})--(\ref{w2t_ii_introd}) for invariants $w_1 = u_{1\xi} - v_{1\eta}$ and $w_2 = u_0 - u_{1\xi} - u_1 v_1$, for the choice $u_1=0$, $v_1 = v$, $u_0 = u$ for which $w_1 = -v_{\eta}$, $w_2 = u$, leads to the well-known system of equations \cite{Konopelchenko_IP_1988}
\begin{gather}
v_{t} = -\kappa_1 v_{\xi \xi} +\kappa_2 v_{\eta \eta} +2\kappa_1 \partial_{\eta}^{-1}u_{\xi \xi} - 2\kappa_2 u_{\eta}+2\kappa_1 vv_{\xi} - 2\kappa_2 v_{\eta} \partial_{\xi}^{-1} v_{\eta}, \label{vt_2DDLW_introd} \\
u_{t} = \kappa_1 u_{\xi \xi} - \kappa_2 u_{\eta \eta} + 2\kappa_1 \bigl(u v\bigr)_{\xi} - 2\kappa_2 \bigl(u \partial_{\xi}^{-1} v_{\eta}\bigr)_{\eta}. \label{ut_2DDLW_introd}
\end{gather}

In terms of variables
\begin{equation}
v = -\frac{q}{2}, \qquad u = \frac{1}{4} (1 + r - q_{\eta}) \label{uv-qr}
\end{equation}
the integrable system of nonlinear equations~(\ref{vt_2DDLW_introd})--(\ref{ut_2DDLW_introd}) takes the form
\begin{gather}
q_t = -\kappa_1 \partial_{\eta}^{-1} r_{\xi \xi} + \kappa_2 r_{\eta} - \frac{\kappa_1}{2} \bigl(q^2\bigr)_{\xi} + \kappa_2 q_{\eta} \partial_{\xi}^{-1} q_{\eta}, \label{qEq2DGDLW} \\
r_t = -\kappa_1 q_{\xi} + \kappa_2 \partial_{\xi}^{-1} q_{\eta \eta} -\kappa_1 q_{\eta \xi \xi} + \kappa_2 q_{\eta \eta \eta} -\kappa_1 \bigl(r q\bigr)_{\xi} + \kappa_2 \bigl(r \partial_{\xi}^{-1} q_{\eta}\bigr)_{\eta}. \label{rEq2DGDLW}
\end{gather}
For the particular value $\kappa_2 = 0$ system of equations~(\ref{qEq2DGDLW})--(\ref{rEq2DGDLW}) reduces to the famous integrable two-dimensional generalization of dispersive long-wave system of equations~\cite{Boiti&Leon&Pempinelli_1987}
\begin{gather}
q_{t \eta} = -\kappa_1 r_{\xi \xi} - \frac{\kappa_1}{2} \bigl(q^2\bigr)_{\xi \eta}, \label{q1Eq2DGDLW} \\
r_{t \xi} = -\kappa_1 \bigl(q r + q + q_{\xi \eta}\bigr)_{\xi \xi}. \label{r1Eq2DGDLW}
\end{gather}
In one-dimensional limit $\xi = \eta$ both systems (\ref{qEq2DGDLW})--(\ref{rEq2DGDLW}) with $\kappa_1-\kappa_2 =1$ and (\ref{q1Eq2DGDLW})--(\ref{r1Eq2DGDLW}) with $\kappa_1 =1$ reduce to the famous dispersive long-wave equation (see, e.g., Broer~\cite {Broer_1975}). It is worthwhile by this reason to name the system (\ref{rho_t_ii_introd})--(\ref{w2t_ii_introd}) as the two-dimensional generalized dispersive long-wave (2DGDLW) system of equations.

In both considered cases of the third- and second-order auxiliary linear problem (\ref{second_auxiliary}) the integrable systems of nonlinear equations~(\ref{rho_t_i_introd})--(\ref{w2t_i_introd}) and (\ref{rho_t_ii_introd})--(\ref{w2t_ii_introd}) have common gauge-transparent structure. They contain correspondingly:
\begin{itemize}
\item gauge-invariant subsystems~(\ref{w2t_i_introd}) and (\ref{w1t_ii_introd})--(\ref{w2t_ii_introd});
\item the equations~(\ref{rho_t_i_introd}) and (\ref{rho_t_ii_introd}) for the pure gauge variable $\rho$ with some terms containing gauge invariants.
\end{itemize}

For the zero values of invariants $w_1 = 0$, $w_2 = 0$ both systems~(\ref{rho_t_i_introd})--(\ref{w2t_i_introd}) and (\ref{rho_t_ii_introd})--(\ref{w2t_ii_introd}) reduce to corresponding linear equations for~$\rho$, respectively,
\begin{equation}
\rho_t = -\kappa_1 \rho_{\xi \xi \xi} - \kappa_2 \rho_{\eta \eta \eta} + v_0 \rho
\end{equation}
and
\begin{equation}
\rho_t = -\kappa_1 \rho_{\xi \xi} - \kappa_2 \rho_{\eta \eta} + v_0 \rho.
\end{equation}

In this paper the NVN (\ref{rho_t_i_introd})--(\ref{w2t_i_introd}) and the 2DGDLW (\ref{rho_t_ii_introd})--(\ref{w2t_ii_introd}) systems of integrable nonlinear equations in different gauges are considered.

It is shown that in some different gauges from~(\ref{rho_t_i_introd})--(\ref{w2t_i_introd}) famous Nizhnik--Veselov--Novikov (NVN) \cite{Nizhnik_1980, Veselov&Novikov_1984} and modified Nizhnik--Veselov--Novikov (mNVN) \cite{Konopelchenko_RMP_1990} equations follow, these equations by Miura-type transformation are connected.

It is shown also that gauge-invariant subsystem (\ref{w1t_ii_introd})--(\ref{w2t_ii_introd}) of the 2DGDLW system~(\ref{rho_t_ii_introd})--(\ref{w2t_ii_introd}) contains in particular, the famous case, integrable two-dimensional generalization of dispersive long-wave system \cite{Boiti&Leon&Pempinelli_1987} of integrable nonlinear equations. In some cases the special gauge 2DGDLW system~(\ref{rho_t_ii_introd})--(\ref{w2t_ii_introd}) reduces to the famous Davey--Stewartson (DS) system \cite{Davey&Stewartson_1974} of nonlinear equations and in another special gauges to new DS-type systems of integrable nonlinear equations, these systems by Miura-type transformation are connected.

The plan of our paper is the following. In sections~\ref{Section_2} and~\ref{Section_3} via the compatibility condition~(\ref{introd_compatibility_condition}) the manifestly gauge-invariant correspondingly integrable NVN system (\ref{rho_t_i_introd})--(\ref{w2t_i_introd}) and the 2DGDLW system~(\ref{rho_t_ii_introd})--(\ref{w2t_ii_introd}) of nonlinear equations are derived. Some special gauges of NVN~(\ref{rho_t_i_introd})--(\ref{w2t_i_introd}) and 2DGDLW~(\ref{rho_t_ii_introd})--(\ref{w2t_ii_introd}) integrable systems of nonlinear equations are considered. Miura-type transformations between solutions of nonlinear equations in different gauges are established.

\section{Manifestly gauge-invariant formulation of NVN system of equations}
\label{Section_2}
\setcounter{equation}{0}

It is instructive to derive integrable nonlinear equations starting from auxiliary linear problems (\ref{first_auxiliary}) and (\ref{second_auxiliary}) in general position, with all nonzero field variables.

Using the compatibility condition (\ref{introd_compatibility_condition}) in the form of Manakov's triad representation~\cite{Manakov_UMN_1976} after some calculations one obtains, equating to zero the coefficients at different degrees of partial derivatives $\partial_{\xi}^n \partial_{\eta}^m$ of the relation $[L_1, L_2] - B L_1 = 0$, the following system of equations for the field variables $u_3$, $v_3$, $u_2$, $v_2$, $\tilde u_1$, $\tilde v_1$, $v_0$ and $u_1$, $v_1$, $u_0$:
\begin{align}
\partial_{\xi}^4: & \quad u_{3\eta} = 0, \qquad \partial_{\eta}^4: \quad v_{3\xi} = 0, \label{xi4_eta4_derivatives} \\
\partial_{\xi}^3 \partial_{\eta}: & \quad u_{3\xi} = 0, \qquad \partial_{\xi} \partial_{\eta}^3: \quad v_{3\eta} = 0, \label{xi3eta1_xi1eta3_derivatives} \\
\partial_{\xi}^3: & \quad u_{3\xi \eta} + u_{2\eta} + u_1 u_{3\xi} - 3u_3 u_{1\xi} + v_1 u_{3\eta} = 0, \label{xi3_derivatives} \\
\partial_{\eta}^3: & \quad v_{3\xi \eta} + v_{2\xi} + v_1 v_{3\eta} - 3v_3 v_{1\eta} + u_1 v_{3\xi} = 0, \label{eta3_derivatives} \\
\partial_{\xi}^2 \partial_{\eta}: & \quad u_{2\xi} - 3u_3 v_{1\xi} = 0, \qquad \partial_{\xi} \partial_{\eta}^2: \quad v_{2\eta} - 3v_3 u_{1\eta} = 0, \label{xi2eta1_xi1eta2_derivatives} \\
\partial_{\xi}^2: & \quad u_{2\xi \eta} + \tilde u_{1\eta} -3u_3 u_{1\xi \xi} - 2u_2 u_{1\xi} + u_1 u_{2\xi} + v_1 u_{2\eta} - 3u_3 u_{0\xi} = 0, \label{xi2_derivatives} \\
\partial_{\eta}^2: & \quad v_{2\xi \eta} + \tilde v_{1\xi} - 3v_3 v_{1\eta \eta} - 2v_2 v_{1\eta} + u_1 v_{2\xi} + v_1 v_{2\eta} - 3v_3 u_{0\eta} = 0, \label{eta2_derivatives} \\
\partial_{\xi \eta}^2: & \quad \tilde u_{1\xi} + \tilde v_{1\eta} - 3u_3 v_{1\xi \xi} - 3v_3 u_{1\eta \eta} - 2u_2 v_{1 \xi} - 2v_2 u_{1\eta} - B = 0, \label{xi1eta1_derivatives} \\
-\partial_{\xi}: & \quad u_{1t} + u_3 u_{1\xi \xi \xi} + v_3 u_{1\eta \eta \eta} + u_2 u_{1\xi \xi} + v_2 u_{1\eta \eta} - v_{0\eta} + \tilde u_1 u_{1\xi} + \tilde v_1 u_{1\eta} \nonumber \\
& \quad {} - u_1 \tilde u_{1\xi} - v_1 \tilde u_{1\eta} - \tilde u_{1\xi \eta} + 3u_3 u_{0\xi \xi} + 2u_2 u_{0\xi} + B u_1 = 0, \label{xi1_derivatives} \\
-\partial_{\eta}: & \quad v_{1t} + u_3 v_{1\xi \xi \xi} + v_3 v_{1\eta \eta \eta} + u_2 v_{1\xi \xi} + v_2 v_{1\eta \eta} - v_{0\xi} + \tilde v_1 v_{1\eta} + \tilde u_1 v_{1\xi} \nonumber \\
& \quad {} - u_1 \tilde v_{1\xi} - v_1 \tilde v_{1\eta} - \tilde v_{1\xi \eta} + 3v_3 u_{0\eta \eta} + 2v_2 u_{0\eta} + B v_1 = 0, \label{eta1_derivatives} \\
-\partial^0: & \quad u_{0t} + u_3 u_{0\xi \xi \xi} + v_3 u_{0 \eta \eta \eta} + u_2 u_{0\xi \xi} + v_2 u_{0\eta \eta} + \tilde u_1 u_{0\xi} + \tilde v_1 u_{0\eta} \nonumber \\
& \quad {} - u_1 v_{0\xi} - v_1 v_{0\eta} - v_{0\xi \eta} + B u_0 = 0. \label{xi0eta0_derivatives}
\end{align}
The system of defining equations~(\ref{xi4_eta4_derivatives})--(\ref{xi0eta0_derivatives}) has recurrent character and allows us to express via~(\ref{xi4_eta4_derivatives})--(\ref{eta2_derivatives}) the field variables $u_3$, $v_3$, $u_2$, $v_2$ and $\tilde u_1$, $\tilde v_1$ of the second auxiliary problem~(\ref{second_auxiliary}) through the field variables $u_1$, $v_1$, $u_0$ of the first auxiliary linear problem~(\ref{first_auxiliary}). The last three equations~(\ref{xi1_derivatives})--(\ref{xi0eta0_derivatives}) represent the integrable system of nonlinear evolution equations for the field variables $u_1,v_1$ and $u_0$.

In the case of the second auxiliary linear problem (\ref{second_auxiliary}) of third order from relations~(\ref{xi4_eta4_derivatives}) and (\ref{xi3eta1_xi1eta3_derivatives}) it follows that the coefficients $u_3$ and $v_3$ are constants,
\begin{equation}
u_3 = \mbox{const} = \kappa_1, \qquad v_3 = \mbox{const} = \kappa_2. \label{u3_v3_subsect2_1}
\end{equation}
Using~(\ref{u3_v3_subsect2_1}) one obtains from the relations~(\ref{xi3_derivatives})--(\ref{xi2eta1_xi1eta2_derivatives}),
\begin{gather}
u_{2\xi} = 3\kappa_1 v_{1\xi}, \qquad v_{2\eta} = 3\kappa_2 u_{1\eta}, \label{u2xi_v2eta_subsect2_1} \\
u_{2\eta} = 3\kappa_1 u_{1\xi}, \qquad v_{2\xi} = 3\kappa_2 v_{1\eta}. \label{u2eta_v2xi_subsect2_1}
\end{gather}
From~(\ref{u2xi_v2eta_subsect2_1})--(\ref{u2eta_v2xi_subsect2_1}) the important relation between field variables $u_1$, $v_1$,
\begin{equation}
u_{1\xi} = v_{1\eta} \label{u1xi=v1eta_subsect2_1}
\end{equation}
and expressions for variables $u_2$ and $v_2$,
\begin{equation}
u_2 = 3\kappa_1 v_1 + \mbox{const}_1, \qquad v_2 = 3\kappa_2 u_1 + \mbox{const}_2 \label{u2_v2_subsect2_1},
\end{equation}
follow. Arising in (\ref{u2_v2_subsect2_1}), for simplicity the constants being equal to zero are chosen below. By the use of~(\ref{xi2_derivatives}) and (\ref{eta2_derivatives}) taking into account (\ref{u3_v3_subsect2_1}), (\ref{u1xi=v1eta_subsect2_1}) and (\ref{u2_v2_subsect2_1}) one derives the expressions for~$\tilde u_1$ and $\tilde v_1$,
\begin{gather}
\tilde u_1 = 3\kappa_1 \partial_{\eta}^{-1} u_{0\xi} - 3\kappa_1 \partial_{\eta}^{-1} (u_1 v_{1\xi}) + \frac{3\kappa_1}{2} v_1^2 + f_1(\xi, t), \label{tilde_u1_subsect2_1} \\
\tilde v_1 = 3\kappa_2 \partial_{\xi}^{-1} u_{0\eta} - 3\kappa_2 \partial_{\xi}^{-1} (v_1 u_{1\eta}) + \frac{3\kappa_2}{2} u_1^2 + g_1(\eta, t), \label{tilde_v1_subsect2_1}
\end{gather}
including as `constants' of integration the arbitrary functions~$f_1(\xi, t)$ and $g_1(\eta, t)$ which for simplicity are chosen below as equal to zero values. Inserting~$\tilde u_1$ and $\tilde v_1$ from~(\ref{tilde_u1_subsect2_1}), (\ref{tilde_v1_subsect2_1}) into~(\ref{xi1eta1_derivatives}) and taking into account (\ref{w2_introd}), (\ref{u3_v3_subsect2_1}), (\ref{u1xi=v1eta_subsect2_1})--(\ref{tilde_v1_subsect2_1}) one derives the expression for the coefficient~$B$,
\begin{align}
B = &-3\kappa_1 v_{1\xi \xi} -3\kappa_2 u_{1\eta \eta} - 3\kappa_1 v_1 v_{1\xi} - 3\kappa_2 u_1 u_{1\eta} + 3\kappa_1 \partial_{\eta}^{-1} u_{0\xi \xi} + 3\kappa_2 \partial_{\xi}^{-1} u_{0\eta \eta} \nonumber \\
{} &- 3\kappa_1 \partial_{\eta}^{-1} \bigl(u_1 v_{1\xi}\bigr)_{\xi} - 3\kappa_2 \partial_{\xi}^{-1} \bigl(v_1 u_{1\eta}\bigr)_{\eta} = 3\kappa_1 \partial_{\eta}^{-1} w_{2\xi \xi} + 3\kappa_2 \partial_{\xi}^{-1} w_{2\eta \eta}. \label{B_subsect2_1}
\end{align}
The last three equations~(\ref{xi1_derivatives})--(\ref{xi0eta0_derivatives}) of the system~(\ref{xi4_eta4_derivatives})--(\ref{xi0eta0_derivatives}) are the evolution equations for the field variables $u_1$, $v_1$ and $u_0$. By the use of~(\ref{w2_introd}), (\ref{u3_v3_subsect2_1}), (\ref{u1xi=v1eta_subsect2_1})--(\ref{B_subsect2_1}) after some calculations (by singling out in some terms the
combination of field variables $ w_2 = u_0 - u_{1\xi} - u_1 v_1$ coinciding with gauge invariant~(\ref{w2_introd})) these equations can be represented in the following convenient form:
\begin{gather}
u_{1t} = -\kappa_1 u_{1\xi \xi \xi} - \kappa_2 u_{1\eta \eta \eta} - \kappa_1 \bigl(v_1^3 + 3v_1 v_{1\xi}\bigr)_{\eta} - \kappa_2 \bigl(u_1^3 + 3u_1 u_{1\eta}\bigr)_{\eta} \nonumber \\
\qquad {} - 3\kappa_1 \bigl(v_1 \partial_{\eta}^{-1} w_{2\xi}\bigr)_{\eta} - 3\kappa_2 \bigl(u_1 \partial_{\xi}^{-1} w_{2\eta}\bigr)_{\eta} + v_{0\eta}, \label{u1t_sect2} \\
v_{1t} = -\kappa_1 v_{1\xi \xi \xi} - \kappa_2 v_{1\eta \eta \eta} - \kappa_1 \bigl(v_1^3 + 3v_1 v_{1\xi}\bigr)_{\xi} - \kappa_2 \bigl(u_1^3 + 3u_1 u_{1\eta}\bigr)_{\xi} \nonumber \\
\qquad {} - 3\kappa_1 \bigl(v_1 \partial_{\eta}^{-1} w_{2\xi}\bigr)_{\xi} - 3\kappa_2 \bigl(u_1 \partial_{\xi}^{-1} w_{2\eta}\bigr)_{\xi} + v_{0\xi}, \label{v1t_sect2} \\
u_{0t} = -\kappa_1 u_{0\xi \xi \xi} - \kappa_2 u_{0\eta \eta \eta} - 3\kappa_1 v_1 u_{0\xi \xi} - 3\kappa_2 u_1 u_{0\eta \eta} - 3\kappa_1 \bigl(v_{1\xi} + v_1^2\bigr) u_{0\xi} - 3\kappa_2 \bigl(u_{1\eta} + u_1^2\bigr) u_{0\eta} \nonumber \\
\qquad {} - 3\kappa_1 \bigl(u_0 \partial_{\eta}^{-1} w_{2\xi}\bigr)_{\xi} - 3\kappa_2 \bigl(u_0 \partial_{\xi}^{-1} w_{2\eta}\bigr)_{\eta} + v_{0\xi \eta} + u_1 v_{0\xi} + v_1 v_{0\eta}. \label{u0t_sect2}
\end{gather}
Remember that in the considered case due to~(\ref{u1xi=v1eta_subsect2_1}) the first invariant $w_1 = u_{1\xi} - v_{1\eta} = 0$ is equal to zero.

Due to the equality~$u_{1\xi} = v_{1\eta}$ one can reduce the set of dependent variables~$u_1$, $v_1$ and $u_0$ in the system~(\ref{u1t_sect2})--(\ref{u0t_sect2}) to two variables~$\rho$, $w_2$ (or equivalently to variables~$\phi = \ln{\rho}$, $w_2$) defined by the relations
\begin{gather}
u_1 \stackrel{\mathrm{def}}{=} \phi_{\eta} = \frac{\rho_{\eta}}{\rho}, \qquad v_1 \stackrel{\mathrm{def}}{=} \phi_{\xi} = \frac{\rho_{\xi}}{\rho}, \label{rho_sect2}\\
w_2 = u_0 - u_{1\xi} - u_1 v_1 = u_0 - \phi_{\xi \eta} - \phi_{\xi} \phi_{\eta} = u_0 - \frac{\rho_{\xi \eta}}{\rho} \label{w_2_sect2}.
\end{gather}
Indeed the insertion of~$u_1 = \phi_{\eta}$ and $v_1 = \phi_{\xi}$ into~(\ref{u1t_sect2}) and (\ref{v1t_sect2}) reduces both these equations to the single one equation
\begin{gather}
\phi_t = -\kappa_1 \phi_{\xi \xi \xi} - \kappa_2 \phi_{\eta \eta \eta} - \kappa_1 (\phi_{\xi})^3 - \kappa_2 (\phi_{\eta})^3 - 3\kappa_1 \phi_{\xi} \phi_{\xi \xi} - 3\kappa_2 \phi_{\eta} \phi_{\eta \eta} \nonumber \\
\qquad {} - 3\kappa_1 \phi_{\xi} \partial_{\eta}^{-1} w_{2\xi} - 3\kappa_2 \phi_{\eta} \partial_{\xi}^{-1} w_{2\eta} + v_0, \label{phiEq_sect2}
\end{gather}
or in terms of variables~$\rho$, $w_2$ to the equation
\begin{equation}
\rho_t = -\kappa_1 \rho_{\xi \xi \xi} - \kappa_2 \rho_{\eta \eta \eta} - 3\kappa_1 \rho_{\xi} \partial_{\eta}^{-1} w_{2\xi} - 3\kappa_2 \rho_{\eta} \partial_{\xi}^{-1} w_{2\eta} + v_0 \rho. \label{rhoEq_sect2}
\end{equation}

One can show also that the exclusion of field variable $v_0$ from the last equation~(\ref{u0t_sect2}) by the use of derivatives~$v_{0\xi}$, $v_{0\eta}$ and $v_{0\xi \eta}$ (calculated from the first two equations~(\ref{u1t_sect2}) and (\ref{v1t_sect2})) leads to the following nonlinear evolution equation for the second invariant~$w_2$:
\begin{equation}
w_{2t} = -\kappa_1 w_{2\xi \xi \xi} - \kappa_2 w_{2\eta \eta \eta} - 3\kappa_1 \bigl(w_2 \partial_{\eta}^{-1} w_{2\xi}\bigr)_{\xi} - 3\kappa_2 \bigl(w_2 \partial_{\xi}^{-1} w_{2\eta}\bigr)_{\eta}. \label{w_2Eq_sect2}
\end{equation}

So by the change of variables~(\ref{rho_sect2}), (\ref{w_2_sect2}) the integrable system of nonlinear equations~(\ref{u1t_sect2})--(\ref{u0t_sect2}) is reduced to the following equivalent integrable system of nonlinear equations:
\begin{gather}
\rho_t = -\kappa_1 \rho_{\xi \xi \xi} - \kappa_2 \rho_{\eta \eta \eta} - 3\kappa_1 \rho_{\xi} \partial_{\eta}^{-1} w_{2\xi} - 3\kappa_2 \rho_{\eta} \partial_{\xi}^{-1} w_{2\eta} + v_0 \rho, \label{rhoEqS_sect2} \\
w_{2t} = -\kappa_1 w_{2\xi \xi \xi} - \kappa_2 w_{2\eta \eta \eta} - 3\kappa_1 \bigl(w_2 \partial_{\eta}^{-1} w_{2\xi}\bigr)_{\xi} - 3\kappa_2 \bigl(w_2 \partial_{\xi}^{-1} w_{2\eta}\bigr)_{\eta}. \label{w_2EqS_sect2}
\end{gather}
Equivalently, in terms of variables~$\phi = \ln{\rho}$ and $w_2$, the system of equations~(\ref{rhoEqS_sect2})--(\ref{w_2EqS_sect2}) takes the form
\begin{gather}
\phi_t = -\kappa_1 \phi_{\xi \xi \xi} - \kappa_2 \phi_{\eta \eta \eta} - \kappa_1 (\phi_{\xi})^3 - \kappa_2 (\phi_{\eta})^3 - 3\kappa_1 \phi_{\xi} \phi_{\xi \xi} - 3\kappa_2 \phi_{\eta} \phi_{\eta \eta} \nonumber \\
\qquad {} - 3\kappa_1 \phi_{\xi} \partial_{\eta}^{-1} w_{2\xi}- 3\kappa_2 \phi_{\eta} \partial_{\xi}^{-1} w_{2\eta} + v_0, \label{phiEqS_sect2}\\
w_{2t} = -\kappa_1 w_{2\xi \xi \xi} - \kappa_2 w_{2\eta \eta \eta} - 3\kappa_1 \bigl(w_2 \partial_{\eta}^{-1} w_{2\xi}\bigr)_{\xi} - 3\kappa_2 \bigl(w_2 \partial_{\xi}^{-1} w_{2\eta}\bigr)_{\eta}. \label{w_2phiEqS_sect2}
\end{gather}
Note that equation~(\ref{w_2EqS_sect2}) (or (\ref{w_2phiEqS_sect2})) for the gauge invariant $w_2$ exactly coincides in form with the famous NVN equation~\cite{Nizhnik_1980, Veselov&Novikov_1984}. Due to this reason it is worthwhile to name the integrable systems~(\ref{rhoEqS_sect2})--(\ref{w_2EqS_sect2}) (or (\ref{phiEqS_sect2})--(\ref{w_2phiEqS_sect2})) as the NVN system of equations.

The NVN system of equations~(\ref{rhoEqS_sect2})--(\ref{w_2EqS_sect2}) (or (\ref{phiEqS_sect2})--(\ref{w_2phiEqS_sect2})) has gauge-transparent structure. It contains:
\begin{itemize}
\item explicitly gauge-invariant subsystem --- equation (\ref{w_2EqS_sect2}) (or (\ref{w_2phiEqS_sect2})) for invariant~$w_2$;
\item equation (\ref{rhoEqS_sect2}) (or (\ref{phiEqS_sect2})) for pure gauge variable~$\rho$ (or~$\phi$) with some terms containing gauge invariant~$w_2$ and field variable~$v_0$ from the second linear auxiliary problem (\ref{second_auxiliary}).
\end{itemize}

Manakov's triad representation~(\ref{introd_compatibility_condition}) for the NVN system of equations~(\ref{rhoEqS_sect2})--(\ref{w_2EqS_sect2}) (or (\ref{phiEqS_sect2})--(\ref{w_2phiEqS_sect2})), due to formulae~(\ref{u3_v3_subsect2_1})--(\ref{B_subsect2_1}) and (\ref{rho_sect2})--(\ref{w_2_sect2}), includes the following operators~$L_1$, $L_2$ of auxiliary linear problems and coefficient~$B(w_2)$:
\begin{gather}
L_1 = \partial_{\xi \eta}^2 + \frac{\rho_{\eta}}{\rho} \partial_{\xi} + \frac{\rho_{\xi}}{\rho} \partial_{\eta} + w_2 + \frac{\rho_{\xi \eta}}{\rho}, \label{L_1mg-i_sect2}\\
L_2 = \partial_t + \kappa_1 \partial_{\xi}^3 + \kappa_2 \partial_{\eta}^3 + 3\kappa_1 \frac{\rho_{\xi}}{\rho} \partial_{\xi}^2 + 3\kappa_2 \frac{\rho_{\eta}}{\rho} \partial_{\eta}^2 + 3\kappa_1 \Bigr(\frac{\rho_{\xi \xi}}{\rho} + \bigl(\partial_{\eta}^{-1} w_{2\xi}\bigr)\Bigl) \partial_{\xi} \nonumber \\
\qquad {} + 3\kappa_2 \Bigl(\frac{\rho_{\eta \eta}}{\rho} + \bigl(\partial_{\xi}^{-1} w_{2\eta}\bigr)\Bigr) \partial_{\eta} + v_0, \label{L_2mg-i_sect2}\\
B(w_2) = 3\kappa_1 \partial_{\eta}^{-1} w_{2\xi \xi} + 3\kappa_2 \partial_{\xi}^{-1} w_{2\eta \eta}. \label{B_mg-i_sect2}
\end{gather}
In the case~$w_2 = 0$ of zero invariant the NVN system of equations~(\ref{rhoEqS_sect2})--(\ref{w_2EqS_sect2}) (or (\ref{phiEqS_sect2})--(\ref{w_2phiEqS_sect2})) reduces to linear equation
\begin{equation}
\rho_t = -\kappa_1 \rho_{\xi \xi \xi} - \kappa_2 \rho_{\eta \eta \eta} + v_0 \rho \label{Linrho_sect2},
\end{equation}
which is integrable by auxiliary linear problems~(\ref{first_auxiliary}) and (\ref{second_auxiliary}) with~$L_1$ and $L_2$ from~(\ref{L_1mg-i_sect2}), (\ref{L_2mg-i_sect2}) under~$w_2 = 0$. The compatibility condition in this case, due to $B(w_2) = 0$, has Lax form $[L_1, L_2] = 0$. In terms of variable~$\phi = \ln{\rho}$ linear equation~(\ref{Linrho_sect2}) looks like Burgers-type equation of the third order
\begin{equation}
\phi_t = -\kappa_1 \phi_{\xi \xi \xi} - \kappa_2 \phi_{\eta \eta \eta} - \kappa_1 (\phi_{\xi})^3 - \kappa_2 (\phi_{\eta})^3 - 3\kappa_1 \phi_{\xi} \phi_{\xi \xi} - 3\kappa_2 \phi_{\eta} \phi_{\eta \eta} + v_0,
\end{equation}
which linearizes by the substitution~$\phi = \ln{\rho}$ and consequently is C-integrable.

Let us denote by~$C \left(\phi, u_0,v_0\right)$ the gauge which corresponds to nonzero field variables~$u_1 = \phi_{\eta}$, $v_1 = \phi_{\xi}$, $u_0$ and $v_0$ of linear problems~(\ref{first_auxiliary}) and (\ref{second_auxiliary}) and consequently to NVN system~(\ref{phiEqS_sect2})--(\ref{w_2phiEqS_sect2}) in general position. Under different gauges from NVN system different integrable nonlinear equations follow, which are gauge-equivalent to each other. The solutions of these equations by some Miura-type transformation are connected.

For example in the gauge $C \left(0, u_0, 0\right)$ the NVN system of equations~(\ref{phiEqS_sect2})--(\ref{w_2phiEqS_sect2}) reduces to the famous NVN equation~\cite{Nizhnik_1980, Veselov&Novikov_1984} for the field variable~$u_0$,
\begin{equation}
u_{0t} = -\kappa_1 u_{0\xi \xi \xi} - \kappa_2 u_{0\eta \eta \eta} - 3\kappa_1 \bigl(u_0 \partial_{\eta}^{-1} u_{0\xi}\bigr)_{\xi} - 3\kappa_2 \bigl(u_0 \partial_{\xi}^{-1} u_{0\eta}\bigr)_{\eta}. \label{NVN_sect2}
\end{equation}
In another gauge $C \left(\phi, 0, v_0\right)$ the NVN system~(\ref{phiEqS_sect2})--(\ref{w_2phiEqS_sect2}) takes the form
\begin{gather}
\phi_t = -\kappa_1 \phi_{\xi \xi \xi} - \kappa_2 \phi_{\eta \eta \eta} - \kappa_1 (\phi_{\xi})^3 - \kappa_2 (\phi_{\eta})^3 + 3\kappa_1 \phi_{\xi} \partial_{\eta}^{-1} \bigl(\phi_{\xi} \phi_{\eta}\bigr)_{\xi} + 3\kappa_2 \phi_{\eta} \partial_{\xi}^{-1} \bigl(\phi_{\xi} \phi_{\eta}\bigr)_{\eta} + v_0, \label{phi1_v_0mNVN} \\
\bigl(\partial_{\xi \eta}^2 + \phi_{\eta} \partial_{\xi} + \phi_{\xi} \partial_{\eta}\bigr) \phi_t = \bigl(\partial_{\xi \eta}^2 + \phi_{\eta} \partial_{\xi} + \phi_{\xi} \partial_{\eta}\bigr) \Bigl[-\kappa_1 \phi_{\xi \xi \xi} - \kappa_2 \phi_{\eta \eta \eta} - \kappa_1 (\phi_{\xi})^3 \nonumber \\
\qquad {} - \kappa_2 (\phi_{\eta})^3 + 3\kappa_1 \phi_{\xi} \partial_{\eta}^{-1} \bigl(\phi_{\xi} \phi_{\eta}\bigr)_{\xi} + 3\kappa_2 \phi_{\eta} \partial_{\xi}^{-1} \bigl(\phi_{\xi} \phi_{\eta}\bigr)_{\eta}\Bigr], \label{phi2_v_0mNVN}
\end{gather}
and consequently to the following system of equations:
\begin{gather}
\phi_t = -\kappa_1 \phi_{\xi \xi \xi} - \kappa_2 \phi_{\eta \eta \eta} - \kappa_1 (\phi_{\xi})^3 - \kappa_2 (\phi_{\eta})^3 + 3\kappa_1 \phi_{\xi} \partial_{\eta}^{-1} \bigl(\phi_{\xi} \phi_{\eta}\bigr)_{\xi} + 3\kappa_2 \phi_{\eta} \partial_{\xi}^{-1} \bigl(\phi_{\xi} \phi_{\eta}\bigr)_{\eta}+v_0, \label{phi_NewmNVNSyst} \\
\bigl(\partial_{\xi \eta}^2 + \phi_{\eta} \partial_{\xi} + \phi_{\xi} \partial_{\eta}\bigr) v_0 = 0 \label{v_0_NewmNVNSyst}
\end{gather}
is equivalent. For $v_0 = 0$ system of equations~(\ref{phi_NewmNVNSyst})--(\ref{v_0_NewmNVNSyst}) reduces to the famous modified Nizhnik--Veselov--Novikov equation
\begin{equation}
\phi_t = -\kappa_1 \phi_{\xi \xi \xi} - \kappa_2 \phi_{\eta \eta \eta} - \kappa_1 (\phi_{\xi})^3 - \kappa_2 (\phi_{\eta})^3 + 3\kappa_1 \phi_{\xi} \partial_{\eta}^{-1} \bigl(\phi_{\xi} \phi_{\eta}\bigr)_{\xi} + 3\kappa_2 \phi_{\eta} \partial_{\xi}^{-1} \bigl(\phi_{\xi} \phi_{\eta}\bigr)_{\eta}, \label{mNVN_sect2}
\end{equation}
which at first in the paper~\cite{Konopelchenko_RMP_1990} of Konopelchenko in a different context was discovered. Let us mention that the considered version (\ref{mNVN_sect2}) of mNVN equation derived in the present paper in the framework of manifestly gauge-invariant description is different from the mNVN equation discovered in the paper~\cite{Bogdanov_1987}.

The new system of equations~(\ref{phi_NewmNVNSyst})--(\ref{v_0_NewmNVNSyst}) can be named as modified NVN (mNVN) system of equations. This system due to (\ref{L_1mg-i_sect2})--(\ref{B_mg-i_sect2}) and to the choice of the gauge $C \left(\phi, 0, v_0\right)$ has the following triad representation (\ref{introd_compatibility_condition}) with triad $(L_1, L_2, B)$:
\begin{gather}
L_1 = \partial_{\xi \eta}^2 + \phi_{\eta} \partial_{\xi} + \phi_{\xi} \partial_{\eta}, \label{L_1mNVNSyst_sect2} \\
L_2 = \partial_t + \kappa_1 \partial_{\xi}^3 + \kappa_2 \partial_{\eta}^3 + 3\kappa_1 \phi_{\xi} \partial_{\xi}^2 + 3\kappa_2 \phi_{\eta} \partial_{\eta}^2 + 3\kappa_1 \Bigr(\phi_{\xi}^2 - \partial_{\eta}^{-1} \bigl(\phi_{\xi} \phi_{\eta}\bigr)_{\xi}\Bigl) \partial_{\xi} \nonumber \\
\qquad {} + 3\kappa_2 \Bigl(\phi_{\eta}^2 - \partial_{\xi}^{-1} \bigl(\phi_{\xi} \phi_{\eta}\bigr)_{\eta}\Bigr) \partial_{\eta} + v_0, \label{L_2mNVNSyst_sect2} \\
B(w_2) = -3\kappa_1 \phi_{\xi \xi \xi} - 3\kappa_2 \phi_{\eta \eta \eta} - 3\kappa_1 \partial_{\eta}^{-1} \bigl(\phi_{\xi} \phi_{\eta}\bigr)_{\xi \xi} - 3\kappa_2 \partial_{\xi}^{-1} \bigl(\phi_{\xi} \phi_{\eta}\bigr)_{\eta \eta}. \label{B_mNVNSyst_sect2}
\end{gather}
The mNVN equation~(\ref{mNVN_sect2}) has triad representation~(\ref{L_1mNVNSyst_sect2})--(\ref{B_mNVNSyst_sect2}) with $v_0 = 0$.

It is evident that the solutions~$u_0$ and $\phi$ of NVN~(\ref{NVN_sect2}) and mNVN~(\ref{mNVN_sect2}) equations via invariant $w_2 = u_0 = -\phi_{\xi \eta} - \phi_{\xi} \phi_{\eta}$ (calculated in different gauges $C \left(0, u_0, 0\right)$ and $C \left(\phi, 0, 0\right)$) by Miura-type transformation
\begin{equation}
u_0 = -\phi_{\xi \eta} - \phi_{\xi} \phi_{\eta}
\end{equation}
are connected. In one-dimensional limit, under $\partial_{\xi} = \partial_{\eta}$, the mNVN equation~(\ref{mNVN_sect2}) reduces to the mKdV equation in potential form
\begin{equation}
\phi_t = -\kappa \, \phi_{\xi \xi \xi} + 2\kappa (\phi_{\xi})^3,
\end{equation}
where $\kappa = \kappa_1 + \kappa_2$. In terms of variable~$v_1 = \phi_{\xi}$ this is mKdV equation
\begin{equation}
v_{1t} = -\kappa \, v_{1\xi \xi \xi} + 6\kappa \, v_1^2 v_{1\xi}.
\end{equation}

\section{Manifestly gauge-invariant formulation of two-dimensional generalization of the dispersive long-wave equations system}
\label{Section_3}
\setcounter{equation}{0}

In the case of second-order linear auxiliary problem~(\ref{second_auxiliary}) the coefficients $u_3,v_3$ in the system of relations~(\ref{xi4_eta4_derivatives})--(\ref{xi0eta0_derivatives}) have zero values $u_3 = v_3 = 0$. The relations~(\ref{xi3_derivatives})--(\ref{xi2eta1_xi1eta2_derivatives}) lead to constant values for the coefficients~$u_2$ and $v_2$,
\begin{equation}
u_2 = \mbox{const} = \kappa_1, \qquad v_2 = \mbox{const} = \kappa_2. \label{u2_v2_ii}
\end{equation}
By integration of relations~(\ref{xi2_derivatives}) and (\ref{eta2_derivatives}) one immediately obtains the expressions for the coefficients~$\tilde u_1$ and $\tilde v_1$,
\begin{equation}
\tilde u_1 = 2\kappa_1 \partial_{\eta}^{-1} u_{1\xi} + f_2(\xi, t), \qquad \tilde v_1 = 2\kappa_2 \partial_{\xi}^{-1} v_{1\eta} + g_2(\eta,t), \label{tilde_u1_tilde_v1_ii}
\end{equation}
where $f_2(\xi, t)$ and $g_2(\eta, t)$ are arbitrary functions which below, for simplicity, chosen equal to zero values. Inserting~(\ref{u2_v2_ii})--(\ref{tilde_u1_tilde_v1_ii}) into~(\ref{xi1eta1_derivatives}) one obtains taking into account~(\ref{w1_introd}) the expression for coefficient~$B$,
\begin{equation}
B = -2\kappa_1 v_{1\xi} - 2\kappa_2 u_{1\eta} + 2\kappa_1 \partial_{\eta}^{-1} u_{1\xi \xi} + 2\kappa_2 \partial_{\xi}^{-1} v_{1\eta \eta}=2\kappa_1\partial_{\eta}^{-1} w_{1\xi}-2\kappa_2 \partial_{\xi}^{-1} w_{1\eta}. \label{B_ii_sect2}
\end{equation}

The last three relations~(\ref{xi1_derivatives})--(\ref{xi0eta0_derivatives}) of the system~(\ref{xi4_eta4_derivatives})--(\ref{xi0eta0_derivatives}) are nonlinear evolution equations for the field variables~$u_1$, $v_1$ and $u_0$. By the use of~(\ref{u2_v2_ii})--(\ref{B_ii_sect2}) after some calculations these equations can be represented (by singling out in some terms the combinations of field variables $w_1=u_{1\xi}-v_{1\eta}$ and $ w_2 = u_0 - u_{1\xi} - u_1 v_1$ coinciding with gauge invariants~(\ref{w2_introd})--(\ref{w1_introd})) in the following convenient form:
\begin{gather}
u_{1t} = -\kappa_1 v_{1\xi \eta} - \kappa_2 u_{1\eta \eta} - 2\kappa_2 u_1 u_{1\eta} - \kappa_1 w_{1\xi} - 2\kappa_1 w_{2\xi} - 2\kappa_1 u_{1\xi} \partial_{\eta}^{-1} u_{1\xi} \nonumber \\
\qquad {} + 2\kappa_2 \bigl(u_1 \partial_{\xi}^{-1} w_1\bigr)_{\eta} + v_{0\eta}, \label{Eqvu1t_subsect2_2} \\
v_{1t} = - \kappa_1 v_{1\xi \xi} - \kappa_2 u_{1\xi \eta} - 2\kappa_1 v_1 v_{1\xi} - \kappa_2 w_{1\eta} - 2\kappa_2 w_{2\eta} - 2\kappa_2 v_{1\eta} \partial_{\xi}^{-1} v_{1\eta} \nonumber \\
\qquad {} - 2\kappa_1 \bigl(v_1 \partial_{\eta}^{-1} w_1\bigr)_{\xi} + v_{0\xi}, \label{Eqvv1t_subsect2_2} \\
u_{0t} = -\kappa_1 u_{0\xi \xi} - \kappa_2 u_{0\eta \eta} - 2\kappa_1 u_{0\xi} v_1 - 2\kappa_2 u_{0\eta} u_1 - 2\kappa_1 \bigl(u_0 \partial_{\eta}^{-1} w_1\bigr)_{\xi} \nonumber \\
\qquad {} + 2\kappa_2 \bigl(u_0 \partial_{\xi}^{-1} w_1\bigr)_{\eta} + v_{0\xi \eta} + u_1 v_{0\xi} + v_1 v_{0\eta}. \label{Eqvu0t_subsect2_2}
\end{gather}

Let us emphasize that the integrable system of nonlinear equations~(\ref{Eqvu1t_subsect2_2})--(\ref{Eqvu0t_subsect2_2}) arises as a compatibility condition of auxiliary linear problems~(\ref{first_auxiliary}) and (\ref{second_auxiliary}) in the form~(\ref{introd_compatibility_condition}) of Manakov's triad representation in the general position. The system contains three evolution equations for the field variables~$u_1$, $v_1$ and $u_0$. These equations include also the field variable~$v_0$ from the second auxiliary linear problem. The presence of these four dependent variables $u_1$, $v_1$, $u_0$ and $v_0$ in system~(\ref{Eqvu1t_subsect2_2})--(\ref{Eqvu0t_subsect2_2}) of three nonlinear equations reflects gauge freedom of auxiliary linear problems (\ref{first_auxiliary}) and (\ref{second_auxiliary}) and the corresponding integrable systems of nonlinear equations. In contrast to the case considered in the previous section, the first invariant $w_1 = u_{1\xi} - v_{1\eta} \ne 0$ is not equal to zero.

One can show that the first two equations~(\ref{Eqvu1t_subsect2_2}) and (\ref{Eqvv1t_subsect2_2}) of last system under change of variables
\begin{gather}
u_1 = \phi_{\eta} = \frac{\rho_{\eta}}{\rho}, \qquad v_1 = -\partial_{\eta}^{-1} w_1 + \phi_{\xi} = -\partial_{\eta}^{-1} w_1 + \frac{\rho_{\xi}}{\rho}, \label{ChangeOfu_1v_1_sect3} \\
w_2 = u_0 - \phi_{\xi \eta} - \phi_{\xi} \phi_{\eta} + \phi_{\eta} \partial_{\eta}^{-1} w_1 = u_0 - \frac{\rho_{\xi \eta}}{\rho} + \frac{\rho_{\eta}}{\rho} \partial_{\eta}^{-1} w_1, \label{ChangeTow_2_sect3}
\end{gather}
reduce to the single one equation of the form
\begin{equation}
\rho_t = -\kappa_1 \rho_{\xi \xi} - \kappa_2 \rho_{\eta \eta} - 2\kappa_1 \rho \partial_{\eta}^{-1} w_{2\xi} + 2\kappa_2 \rho_{\eta} \partial_{\xi}^{-1} w_1 + v_0 \rho, \label{rho_t2DGDLW_sect3}
\end{equation}
or in terms of variable $\phi = \ln{\rho}$ to the equation
\begin{equation}
\phi_t = -\kappa_1 \phi_{\xi \xi} - \kappa_2 \phi_{\eta \eta} - \kappa_1 (\phi_{\xi})^2 - \kappa_2 (\phi_{\eta})^2 - 2\kappa_1 \partial_{\eta}^{-1} w_{2\xi} + 2\kappa_2 \phi_{\eta} \partial_{\xi}^{-1} w_1 + v_0. \label{phi_t2DGDLW_sect3}
\end{equation}

The condition of equality of mixture derivatives~$v_{0\xi \eta}$ and $v_{0\eta \xi}$, calculated from~(\ref{Eqvu1t_subsect2_2}) and (\ref{Eqvv1t_subsect2_2}), leads to the following nonlinear evolution equation in terms of gauge invariants~$w_1$ and $w_2$,
\begin{equation}
w_{1t} = -\kappa_1 w_{1\xi \xi} + \kappa_2 w_{2\eta \eta} - 2\kappa_1 w_{2\xi \xi} + 2\kappa_2 w_{2\eta \eta} - 2\kappa_1 \bigl(w_1 \partial_{\eta}^{-1} w_1\bigr)_{\xi} + 2\kappa_2 \bigl(w_1 \partial_{\xi}^{-1} w_1\bigr)_{\eta}. \label{w_1t_sect3}
\end{equation}
One can show also that the exclusion of free field variable~$v_0$ from the last equation~(\ref{Eqvu0t_subsect2_2}) by the use of derivatives~$v_{0\xi}$, $v_{0\eta}$ and $v_{0\xi \eta}$, calculated from the first two equations~(\ref{Eqvu1t_subsect2_2}) and (\ref{Eqvv1t_subsect2_2}), leads to another evolution equation in terms of invariants~$w_1$ and $w_2$,
\begin{equation}
w_{2t} = \kappa_1 w_{2\xi \xi} - \kappa_2 w_{2\eta \eta} - 2\kappa_1 \bigl(w_2 \partial_{\eta}^{-1} w_1\bigr)_{\xi} + 2\kappa_2 \bigl(w_2 \partial_{\xi}^{-1} w_1\bigr)_{\eta}. \label{w_2t_sect3}
\end{equation}

So by the change of variables~(\ref{ChangeOfu_1v_1_sect3}), (\ref{ChangeTow_2_sect3}) the integrable system~(\ref{Eqvu1t_subsect2_2})--(\ref{Eqvu0t_subsect2_2}) of nonlinear equations of second order is reduced to the following equivalent integrable system of nonlinear equations:
\begin{gather}
\rho_t = -\kappa_1 \rho_{\xi \xi} - \kappa_2 \rho_{\eta \eta} - 2\kappa_1 \rho \partial_{\eta}^{-1} w_{2\xi} + 2\kappa_2 \rho_{\eta} \partial_{\xi}^{-1} w_1 + v_0 \rho, \label{Eq_rho_sect3} \\
w_{1t} = -\kappa_1 w_{1\xi \xi} + \kappa_2 w_{1\eta \eta} - 2\kappa_1 w_{2\xi \xi} + 2\kappa_2 w_{2\eta \eta} - 2\kappa_1 \bigl(w_1 \partial_{\eta}^{-1} w_1\bigr)_{\xi} + 2\kappa_2 \bigl(w_1 \partial_{\xi}^{-1} w_1\bigr)_{\eta}, \label{Eq_w_1_sect3} \\
w_{2t} = \kappa_1 w_{2\xi \xi} - \kappa_2 w_{2\eta \eta} - 2\kappa_1 \bigl(w_2 \partial_{\eta}^{-1} w_1\bigr)_{\xi} + 2\kappa_2 \bigl(w_2 \partial_{\xi}^{-1} w_1\bigr)_{\eta}. \label{Eq_w_2_sect3}
\end{gather}
In terms of variables $\phi = \ln{\rho}$, $w_1$ and $w_2$ the integrable system~(\ref{Eq_rho_sect3})--(\ref{Eq_w_2_sect3}) takes the form
\begin{gather}
\phi_t = -\kappa_1 \phi_{\xi \xi} - \kappa_2 \phi_{\eta \eta} - \kappa_1 (\phi_{\xi})^2 - \kappa_2 (\phi_{\eta})^2 - 2\kappa_1 \partial_{\eta}^{-1} w_{2\xi} + 2\kappa_2 \phi_{\eta} \partial_{\xi}^{-1} w_1 + v_0, \label{phi_w_1W_2_sect3} \\
w_{1t} = -\kappa_1 w_{1\xi \xi} + \kappa_2 w_{1\eta \eta} - 2\kappa_1 w_{2\xi \xi} + 2\kappa_2 w_{2\eta \eta} - 2\kappa_1 \bigl(w_1 \partial_{\eta}^{-1} w_1\bigr)_{\xi} + 2\kappa_2 \bigl(w_1 \partial_{\xi}^{-1} w_1\bigr)_{\eta}, \label{w_1t_w_2_sect3} \\
w_{2t} = \kappa_1 w_{2\xi \xi} - \kappa_2 w_{2\eta \eta} - 2\kappa_1 \bigl(w_2 \partial_{\eta}^{-1} w_1\bigr)_{\xi} + 2\kappa_2 \bigl(w_2 \partial_{\xi}^{-1} w_1\bigr)_{\eta}. \label{w_2t_w_1_sect3}
\end{gather}
In terms of variables $\phi = \ln{\rho}$, $w_2$ and $\widetilde w_2 = w_2 + w_1$ the integrable system~(\ref{Eq_rho_sect3})--(\ref{Eq_w_2_sect3}) converts into more symmetrical form
\begin{gather}
\phi_t = -\kappa_1 \phi_{\xi \xi} - \kappa_2 \phi_{\eta \eta} - \kappa_1 (\phi_{\xi})^2 - \kappa_2 (\phi_{\eta})^2 - 2\kappa_1 \partial_{\eta}^{-1} w_{2\xi} + 2\kappa_2 \phi_{\eta} \partial_{\xi}^{-1} w_1 + v_0, \label{phi_tildew_sect3} \\
w_{2t} = \kappa_1 w_{2\xi \xi} - \kappa_2 w_{2\eta \eta} - 2\kappa_1 \bigl(w_2 \partial_{\eta}^{-1} (\widetilde w_2 - w_2)\bigr)_{\xi} + 2\kappa_2 \bigl(w_2 \partial_{\xi}^{-1} (\widetilde w_2 - w_2)\bigr)_{\eta}, \label{w_2phi_tildew_sect3} \\
\widetilde w_{2t} = - \kappa_1 \widetilde w_{2\xi \xi} + \kappa_2 \widetilde w_{2\eta \eta} - 2\kappa_1 \bigl(\widetilde w_2 \partial_{\eta}^{-1} (\widetilde w_2 - w_2)\bigr)_{\xi} + 2\kappa_2 \bigl(\widetilde w_2 \partial_{\xi}^{-1} (\widetilde w_2 - w_2)\bigr)_{\eta}. \label{widetilde_wphi_tildew_sect3}
\end{gather}

Remember for convenience that due to (\ref{w2_introd})--(\ref{ro_introd}) in equivalent to each other systems of nonlinear equations~(\ref{Eq_rho_sect3})--(\ref{Eq_w_2_sect3}), (\ref{phi_w_1W_2_sect3})--(\ref{w_2t_w_1_sect3}) and (\ref{phi_tildew_sect3})--(\ref{widetilde_wphi_tildew_sect3}) the variables $\phi = \ln{\rho}$, $w_1$, $w_2$ and $\widetilde w_2$ are connected with the field variables~$u_1$, $v_1$, $u_0$ of the linear problem~(\ref{first_auxiliary}) by the formulae
\begin{gather}
u_1 = \frac{\rho_{\eta}}{\rho} = \phi_{\eta}, \qquad v_1 = \frac{\rho_{\xi}}{\rho} - \partial_{\eta}^{-1} w_1 = \phi_{\xi} - \partial_{\eta}^{-1} w_1, \qquad w_1 = u_{1\xi} - v_{1\eta}, \label{Convu_1v_1_sect_3} \\
w_2 = u_0 - u_{1\xi} - u_1 v_1 = u_0 - \phi_{\xi \eta} - \phi_{\eta} \phi_{\xi} + \phi_{\eta} \partial_{\eta}^{-1} w_1 = u_0 - \frac{\rho_{\xi \eta}}{\rho} + \frac{\rho_{\eta}}{\rho} \partial_{\eta}^{-1} w_1, \label{Conv_w_2sect_3} \\
\widetilde w_2 = w_2 + w_1. \label{Conv_widetildew_2sect_3}
\end{gather}

Integrable system of nonlinear equations~(\ref{Eq_rho_sect3})--(\ref{Eq_w_2_sect3}) (and analogously equivalent systems (\ref{phi_w_1W_2_sect3})--(\ref{w_2t_w_1_sect3}) or (\ref{phi_tildew_sect3})--(\ref{widetilde_wphi_tildew_sect3})) for the choice of variables
\begin{equation}
\rho = 1; \qquad u_1 = 0, \; v_1 = v, \; u_0 = u; \qquad v_0 = 2\kappa_1 \partial_{\eta}^{-1} w_{2\xi} \label{DefOf2DGDLW}
\end{equation}
for which $w_1 = -v_{\eta}$, $w_2 = u$, reduces to known system of equations
\begin{gather}
v_{t} = -\kappa_1 v_{\xi \xi} +\kappa_2 v_{\eta \eta} - 2\kappa_2 u_{\eta}+2\kappa_1 vv_{\xi} +2\kappa_1 \partial_{\eta}^{-1}u_{\xi \xi}- 2\kappa_2 v_{\eta} \partial_{\xi}^{-1} v_{\eta}, \label{Defvt_2DDLW_sect3} \\
u_{t} = \kappa_1 u_{\xi \xi} - \kappa_2 u_{\eta \eta} + 2\kappa_1 \bigl(u v\bigr)_{\xi} - 2\kappa_2 \bigl(u \partial_{\xi}^{-1} v_{\eta}\bigr)_{\eta}, \label{Defut_2DDLW_sect3}
\end{gather}
derived in different context by Konopelchenko~\cite{Konopelchenko_IP_1988}.

For the particular values $\kappa_1 = 1$ and $\kappa_2 = 0$, system of equations (\ref{Defvt_2DDLW_sect3})--(\ref{Defut_2DDLW_sect3}) reduces to famous integrable two-dimensional generalization of dispersive long-wave system of equations
\begin{gather}
v_{t \eta} = -v_{\xi \xi \eta} + 2 u_{\xi \xi} + \bigl(v^2\bigr)_{\xi \eta}, \label{vt_2DGDLW_introd} \\
u_t = u_{\xi \xi} + 2\bigl(u v\bigr)_{\xi}, \label{Defut_2DGDLW_sect3}
\end{gather}
discovered by Boiti, Leon and Pempinelli~\cite{Boiti&Leon&Pempinelli_1987}. It is interesting to note that in a different context the system of equations (\ref{w_2phi_tildew_sect3})--(\ref{widetilde_wphi_tildew_sect3}) for Laplace invariants $h = w_2$ and $k = \widetilde w_2$ in the case $\kappa_1 = 1$, $\kappa_2 = 0$ in the paper of Weiss~\cite{Weiss_1991} was considered. By this reason and due to the remarks in section~\ref{Section_1} (see (\ref{rho_t_ii_introd})--(\ref{r1Eq2DGDLW}) and discussion therein) it is worthwhile to name the integrable system of nonlinear equations~(\ref{Eq_rho_sect3})--(\ref{Eq_w_2_sect3}) (and analogously equivalent systems (\ref{phi_w_1W_2_sect3})--(\ref{w_2t_w_1_sect3}) or (\ref{phi_tildew_sect3})--(\ref{widetilde_wphi_tildew_sect3})) as a two-dimensional generalization of dispersive long-wave (2DGDLW) system of equations.

All considered equivalent to each other, 2DGDLW integrable systems of nonlinear equations~(\ref{Eq_rho_sect3})--(\ref{Eq_w_2_sect3}), (\ref{phi_w_1W_2_sect3})--(\ref{w_2t_w_1_sect3}) and (\ref{phi_tildew_sect3})--(\ref{widetilde_wphi_tildew_sect3}) have a common gauge-transparent structure:
\begin{itemize}
\item they contain explicitly gauge-invariant subsystems (\ref{Eq_w_1_sect3})--(\ref{Eq_w_2_sect3}), (\ref{w_1t_w_2_sect3})--(\ref{w_2t_w_1_sect3}) of nonlinear equations for gauge invariants~$w_1$ and $w_2$ (or equivalently subsystem (\ref{w_2phi_tildew_sect3})--(\ref{widetilde_wphi_tildew_sect3}) for gauge invariants~$w_2$ and $\widetilde w_2$);
\item they include equation (\ref{Eq_rho_sect3}) for pure gauge variable~$\rho$ (or equation (\ref{phi_w_1W_2_sect3}) for variable $\phi = \ln{\rho}$) (with simple rule of gauge transformation $\rho \rightarrow \rho' = g \rho$) with additional terms containing gauge invariants and field variable~$v_0$.
\end{itemize}

Such structure of 2DGDLW systems reflects existing gauge freedom in auxiliary linear problems~(\ref{first_auxiliary}) and (\ref{second_auxiliary}).

Due to formulae~(\ref{w2_introd}), (\ref{w1_introd}) and (\ref{u2_v2_ii})--(\ref{B_ii_sect2}) 2DGDLW system~(\ref{Eq_rho_sect3})--(\ref{Eq_w_2_sect3}) has triad representation $[L_1, L_2] = B(w_1) L_1$ with operators~$L_1$, $L_2$ and coefficient~$B(w_1)$ of the following forms:
\begin{gather}
L_1 = \partial_{\xi \eta}^2 + \frac{\rho_{\eta}}{\rho} \partial_{\xi} + \Bigl(\frac{\rho_{\xi}}{\rho} - \bigl(\partial_{\eta}^{-1} w_1\bigr)\Bigr) \partial_{\eta} + w_2 + \frac{\rho_{\xi \eta}}{\rho} - \frac{\rho_{\eta}}{\rho} \, \partial_{\eta}^{-1} w_1, \label{L_1MGInv_sect3} \\
L_2 = \partial_t + \kappa_1 \partial_{\xi}^2 + \kappa_2 \partial_{\eta}^2 + 2\kappa_1 \frac{\rho_{\xi}}{\rho} \partial_{\xi} + 2\kappa_2 \Bigl(\frac{\rho_{\eta}}{\rho} - \bigl(\partial_{\xi}^{-1} w_1\bigr)\Bigr) \partial_{\eta} + v_0, \label{L_2MGInv_sect3} \\
B(w_1) = 2\kappa_1 \partial_{\eta}^{-1} w_{1\xi} - 2\kappa_2 \partial_{\xi}^{-1} w_{1\eta}. \label{BMGInv_sect3}
\end{gather}

Let us consider some particular gauges of established 2DGDLW systems of equations equations~(\ref{Eq_rho_sect3})--(\ref{Eq_w_2_sect3}), (\ref{phi_w_1W_2_sect3})--(\ref{w_2t_w_1_sect3}) and (\ref{phi_tildew_sect3})--(\ref{widetilde_wphi_tildew_sect3}). It is convenient to denote the gauge in general position by the symbol $C \left(u_1, v_1, u_0\right)$.

In the gauge $C \left(u_1 = \phi_{\eta}, v_1 = \phi_{\xi}, u_0 = \phi_{\xi \eta} + \phi_{\xi} \phi_{\eta}\right)$ which due to~(\ref{w2_introd})--(\ref{w1_introd}) corresponds to zero values of invariants~$w_1$ and $w_2$
\begin{equation}
w_1 = u_{1\xi} - v_{1\eta} = 0, \qquad w_2 = u_0 - u_{1\xi} - u_1 v_1 = 0, \qquad \widetilde w_2 = 0,
\end{equation}
the 2DGDLW system of equations (\ref{phi_tildew_sect3})--(\ref{widetilde_wphi_tildew_sect3}) reduces to two-dimensional Burgers equation in potential form
\begin{equation}
\phi_t = -\kappa_1 \phi_{\xi \xi} - \kappa_2 \phi_{\eta \eta} - \kappa_1 (\phi_{\xi})^2 - \kappa_2 (\phi_{\eta})^2 + v_0, \label{phiBurgers_sect3}
\end{equation}
or in terms of variable~$\rho$ connected with~$\phi$ by Hopf-Cole transformation $\phi = \ln{\rho}$, to linear diffusion equation
\begin{equation}
\rho_t = -\kappa_1 \rho_{\xi \xi} - \kappa_2 \rho_{\eta \eta}+v_0\rho. \label{rhoBurgers_sect3}
\end{equation}
Equation~(\ref{phiBurgers_sect3}) (or (\ref{rhoBurgers_sect3})) due to our construction is a compatibility condition in Lax form
\begin{equation}
[L_1, L_2] = B(w_1) L_1 \equiv 0
\end{equation}
of linear problems~(\ref{first_auxiliary}) and (\ref{second_auxiliary}) with operators~$L_1$, $L_2$ given by~(\ref{L_1MGInv_sect3}), (\ref{L_2MGInv_sect3}) under substitution $w_1 = w_2 = 0$.

In another simple gauge $C \left(u_1 = \phi_{\eta}, v_1 = 0, u_0 = 0\right)$ corresponding due to~(\ref{Convu_1v_1_sect_3})--(\ref{Conv_widetildew_2sect_3}) to the invariants
\begin{equation}
w_1 = \phi_{\xi \eta}, \qquad w_2 = -\phi_{\xi \eta}, \qquad \widetilde w_2 = 0,
\end{equation}
the 2DGDLW system of equations (\ref{phi_tildew_sect3})--(\ref{widetilde_wphi_tildew_sect3}) for the choice~$v_0 = 0$ again reduces to the single equation of Burgers type in potential form
\begin{equation}
\phi_t = \kappa_1 \phi_{\xi \xi} - \kappa_2 \phi_{\eta \eta} - \kappa_1 (\phi_{\xi})^2 + \kappa_2 (\phi_{\eta})^2.
\end{equation}
This equation linearizes by Hopf-Cole transformation $\phi = -\ln{\rho}$ to corresponding linear equation
\begin{equation}
\rho_t = \kappa_1 \rho_{\xi \xi} - \kappa_2 \rho_{\eta \eta}.
\end{equation}

In the less trivial gauge $C \left(u_1 = 0, v_1 = -q_{\xi} / q, u_0 = p \, q\right)$ the invariants~$w_1$, $w_2$ and $\widetilde w_2$ due to (\ref{Convu_1v_1_sect_3})--(\ref{Conv_widetildew_2sect_3}) are given by the following expressions:
\begin{equation}
w_1 = \bigl(\ln{q}\bigr)_{\xi \eta}, \qquad w_2 = u_0 = p \, q, \qquad \widetilde w_2 = p \, q + \bigl(\ln{q}\bigr)_{\xi \eta},
\end{equation}
the variable~$\rho$ due to~(\ref{Convu_1v_1_sect_3}) has constant value, consequently the variable $\phi=0$. In this case due to~(\ref{phi_tildew_sect3})
\begin{equation}
v_0 = 2\kappa_1 \partial_{\eta}^{-1} w_{2\xi} = 2\kappa_1 \partial_{\eta}^{-1} \bigl(p \, q\bigr)_{\xi}
\end{equation}
and from the 2DGDLW system of equations~(\ref{phi_tildew_sect3})--(\ref{widetilde_wphi_tildew_sect3}) one obtains after some calculations the famous DS system of equations~\cite{Davey&Stewartson_1974} for the field variables~$p$ and $q$,
\begin{gather}
p_t = \kappa_1 p_{\xi \xi} - \kappa_2 p_{\eta \eta} + 2\kappa_1 p \, \partial_{\eta}^{-1} \bigl(p \, q\bigr)_{\xi} - 2\kappa_2 p \, \partial_{\xi}^{-1} \bigl(p \, q\bigr)_{\eta}, \label{Eq_p_Famous_DS}\\
q_t = -\kappa_1 q_{\xi \xi} + \kappa_2 q_{\eta \eta} - 2\kappa_1 q \, \partial_{\eta}^{-1} \bigl(p \, q\bigr)_{\xi} + 2\kappa_2 q \, \partial_{\xi}^{-1} \bigl(p \, q\bigr)_{\eta}. \label{Eq_q_Famous_DS}
\end{gather}

One can consider also the gauge $C \left(u_1 = p_{\eta}, v_1 = q_{\xi}, u_0 = p_{\eta} q_{\xi}\right)$ in which due to~(\ref{Convu_1v_1_sect_3})--(\ref{Conv_widetildew_2sect_3}) the invariants have the following expressions through~$q$ and $p$:
\begin{equation}
w_1 = p_{\xi \eta} - q_{\xi \eta}, \qquad w_2 = -p_{\xi \eta}, \qquad \widetilde w_2 = -q_{\xi \eta}. \label{2DGLW_DS2_sect3}
\end{equation}
Substitution of~$w_1$, $w_2$ and $\widetilde w_2$ from~(\ref{2DGLW_DS2_sect3}) into the system~(\ref{phi_tildew_sect3})--(\ref{widetilde_wphi_tildew_sect3}) leads to the following three equations for~$p$ and $q$. From equation~(\ref{phi_tildew_sect3}) for~$\phi \equiv p$ one obtains
\begin{equation}
p_t = \kappa_1 p_{\xi \xi} - \kappa_2 p_{\eta \eta} - \kappa_1 (p_{\xi})^2 + \kappa_2 (p_{\eta})^2 - 2\kappa_2 p_{\eta} q_{\eta} + v_0. \label{pEq1_DS2_sect3}
\end{equation}
Equations~(\ref{w_2phi_tildew_sect3}) and (\ref{widetilde_wphi_tildew_sect3}) for~$w_2$ and $\widetilde w_2$ in terms of variables~$p$, $q$ take the forms
\begin{gather}
p_t = \kappa_1 p_{\xi \xi} - \kappa_2 p_{\eta \eta} - \kappa_1 (p_{\xi})^2 + \kappa_2 (p_{\eta})^2 + 2\kappa_1 \partial_{\eta}^{-1} \bigl(p_{\xi \eta} q_{\xi}\bigr) - 2\kappa_2 \partial_{\xi}^{-1} \bigl(p_{\xi \eta} q_{\eta}\bigr), \label{pEq2_DS2_sect3} \\
q_t = -\kappa_1 q_{\xi \xi} + \kappa_2 q_{\eta \eta} + \kappa_1 (q_{\xi})^2 - \kappa_2 (q_{\eta})^2 - 2\kappa_1 \partial_{\eta}^{-1} \bigl(q_{\xi \eta} p_{\xi}\bigr) + 2\kappa_2 \partial_{\xi}^{-1} \bigl(q_{\xi \eta} p_{\eta}\bigr). \label{qEq_DS2-sect3}
\end{gather}
Equations~(\ref{pEq1_DS2_sect3}) and (\ref{pEq2_DS2_sect3}) are compatible for the choice of~$v_0$ given by the formula
\begin{equation}
v_0 = 2\kappa_1 \partial_{\eta}^{-1} \bigl(p_{\xi \eta} q_{\xi}\bigr) + 2\kappa_2 \partial_{\xi}^{-1} \bigl(q_{\xi \eta} p_{\eta}\bigr), \label{v_0DS2_sect3}
\end{equation}
and the system of three equations~(\ref{pEq1_DS2_sect3})--(\ref{qEq_DS2-sect3}) reduces to system of two equations~(\ref{pEq2_DS2_sect3})--(\ref{qEq_DS2-sect3}) containing in nonlocal terms derivatives $p_{\xi \eta} q_{\xi}$, $p_{\xi \eta} q_{\eta}$, etc.

Analogously in the gauge $C \left(u_1 = p_{\eta}, v_1 = q_{\xi}, u_0 = 0\right)$ it follows for~$w_1$, $w_2$ and $\widetilde w_2$ due to~(\ref{Convu_1v_1_sect_3})--(\ref{Conv_widetildew_2sect_3})
\begin{equation}
w_1 = p_{\xi \eta} - q_{\xi \eta}, \qquad w_2 = -p_{\xi \eta} - p_{\eta} q_{\xi}, \qquad \widetilde w_2 = -q_{\xi \eta} - p_{\eta} q_{\xi}. \label{2DGDLW_DS3}
\end{equation}
Equation~(\ref{phi_w_1W_2_sect3}) for $\phi \equiv p$ via~(\ref{2DGDLW_DS3}) takes the form
\begin{equation}
p_t = \kappa_1 p_{\xi \xi} - \kappa_2 p_{\eta \eta} - \kappa_1 (p_{\xi})^2 + \kappa_2 (p_{\eta})^2 - 2\kappa_2 p_{\eta} q_{\eta} + 2\kappa_1 \partial_{\eta}^{-1} \bigl(p_{\eta} q_{\xi}\bigr)_{\xi} + v_0. \label{pEq1_DS3_sect3}
\end{equation}
Equation~(\ref{w_1t_w_2_sect3}) via substitutions from~(\ref{2DGDLW_DS3}) transforms to the form
\begin{gather}
p_t - q_t = \kappa_1 \bigl(p + q\bigr)_{\xi \xi} - \kappa_2 \bigl(p + q\bigr)_{\eta \eta} - \kappa_1 (p_{\xi} - q_{\xi})^2 + \kappa_2 (p_{\eta} - q_{\eta})^2 \nonumber \\
\qquad {} + 2\kappa_1 \partial_{\eta}^{-1} \bigl(p_{\eta} q_{\xi}\bigr)_{\xi} - 2\kappa_2 \partial_{\xi}^{-1} \bigl(p_{\eta} q_{\xi}\bigr)_{\eta}. \label{pEq2_DS3_sect3}
\end{gather}
By substraction of equation~(\ref{pEq2_DS3_sect3}) from equation~(\ref{pEq1_DS3_sect3}) one obtains the evolution equation for~$q$:
\begin{equation}
q_t = -\kappa_1 q_{\xi \xi} + \kappa_2 q_{\eta \eta} + \kappa_1 (q_{\xi})^2 - \kappa_2 (q_{\eta})^2 - 2\kappa_1 p_{\xi} q_{\xi} + 2\kappa_2 \partial_{\xi}^{-1} \bigl(p_{\eta} q_{\xi}\bigr)_{\eta}+v_0. \label{qEq_DS3_sect3}
\end{equation}
Equation (\ref{w_2t_w_1_sect3}) for the invariant~$w_2$ due to~(\ref{2DGDLW_DS3}) in terms of variables~$p$, $q$ is
\begin{gather}
\bigl(p_{\xi \eta} + p_{\eta} q_{\xi}\bigr)_t = \kappa_1 \bigl(p_{\xi \eta} + p_{\eta} q_{\xi}\bigr)_{\xi \xi} - \kappa_2 \bigl(p_{\xi \eta} + p_{\eta} q_{\xi}\bigr)_{\eta \eta} \nonumber \\
\qquad {} - 2\kappa_1 \bigl((p_{\xi \eta} + p_{\eta} q_{\xi}) (p_{\xi} - q_{\xi})\bigr)_{\xi} + 2\kappa_2 \bigl((p_{\xi \eta} + p_{\eta} q_{\xi}) (p_{\eta} - q_{\eta})\bigr)_{\eta}. \label{pqEq_DS3_sect3}
\end{gather}
Equations~(\ref{pEq1_DS3_sect3}), (\ref{qEq_DS3_sect3}) and (\ref{pqEq_DS3_sect3}) are compatible with each other if the field variable~$v_0$ satisfies the equation
\begin{equation}
v_{0\xi \eta} + p_{\eta} v_{0\xi} + q_{\xi} v_{0\eta} = 0.
\end{equation}
For the simple choice~$v_0 \equiv 0$ one obtains from the system of the three equations~(\ref{pEq1_DS3_sect3}), (\ref{qEq_DS3_sect3}) and (\ref{pqEq_DS3_sect3}) the following equivalent system of two equations:
\begin{gather}
p_t = \kappa_1 p_{\xi \xi} - \kappa_2 p_{\eta \eta} - \kappa_1 (p_{\xi})^2 + \kappa_2 (p_{\eta})^2 - 2\kappa_2p_{\eta} q_{\eta} + 2\kappa_1 \partial_{\eta}^{-1} \bigl(p_{\eta} q_{\xi}\bigr)_{\xi}, \label{Eq_p_DS4_sect3} \\
q_t = -\kappa_1 q_{\xi \xi} + \kappa_2 q_{\eta \eta} + \kappa_1 (q_{\xi})^2 - \kappa_2 (q_{\eta})^2 - 2\kappa_1 p_{\xi} q_{\xi} + 2\kappa_2 \partial_{\xi}^{-1} \bigl(p_{\eta} q_{\xi}\bigr)_{\eta}. \label{Ep_q_DS4_sect3}
\end{gather}
At first this system of equations has been derived in another context in the paper of Konopelchenko~\cite{Konopelchenko_IP_1988}.

In conclusion, let us derive Miura-type transformations between different systems of DS-type equations of second order obtained in this section in different gauges. For convenience let us denote by capital letters $P \equiv p$, $Q \equiv q$ the solutions of the DS famous system~(\ref{Eq_p_Famous_DS})--(\ref{Eq_q_Famous_DS}) of equations. By the use of invariants~$w_1$ and $w_2$ one obtains the following relations between variables ($P \equiv p$, $Q \equiv q$) of DS system~(\ref{Eq_p_Famous_DS})--(\ref{Eq_q_Famous_DS}) and variables~$p$, $q$ of the system~(\ref{pEq2_DS2_sect3})--(\ref{qEq_DS2-sect3}),
\begin{equation}
w_1 = \bigl(\ln{Q}\bigr)_{\xi \eta} = p_{\xi \eta} - q_{\xi \eta}, \qquad w_2 = P Q = -p_{\xi \eta}. \label{InvMiura_TypeTr1}
\end{equation}
One derives from~(\ref{InvMiura_TypeTr1}),
\begin{equation}
Q = e^{p - q}, \qquad P = -p_{\xi \eta} \, e^{q - p}. \label{DSMiura_TypeTr1}
\end{equation}

Quite analogously for the pair of DS systems~(\ref{Eq_p_Famous_DS})--(\ref{Eq_q_Famous_DS}) and~(\ref{Eq_p_DS4_sect3})--(\ref{Ep_q_DS4_sect3}) one has
\begin{equation}
w_1 = \bigl(\ln{Q}\bigr)_{\xi \eta} = p_{\xi \eta} - q_{\xi \eta}, \qquad w_2 = P Q = -p_{\xi \eta} - p_{\eta} q_{\xi}. \label{InvMiura_TypeTr2}
\end{equation}
One obtains from~(\ref{InvMiura_TypeTr2}),
\begin{equation}
Q = e^{p - q}, \qquad P = -\bigl(p_{\xi \eta} + p_{\eta} q_{\xi}\bigr) e^{q - p}. \label{DSMiura_TypeTr2}
\end{equation}

Transformations~(\ref{DSMiura_TypeTr1}) and (\ref{DSMiura_TypeTr2}) allow us to obtain solutions of the famous DS system of equations~(\ref{Eq_p_Famous_DS})--(\ref{Eq_q_Famous_DS}) from the systems of equations~(\ref{pEq2_DS2_sect3})--(\ref{qEq_DS2-sect3}) and~(\ref{Eq_p_DS4_sect3})--(\ref{Ep_q_DS4_sect3}), these transformations are Miura-type transformations being gauge-equivalent to other DS-type systems of equations of second order.

\section{Conclusion}

In conclusion let us underline once again that ideas of gauge invariance now are in common use in the theory of integrable nonlinear evolution equations. There are known attempts to develop invariant description of some nonlinear integrable equations considered in the present paper by the use of matrix linear auxiliary problems. This was done for example in the paper~\cite{Yilmaz&Athorne_2002} for the Nizhnik--Veselov--Novikov and Davey--Stewartson equations in the framework of the classical invariant theory of second-order linear partial differential equations.

Matrix linear auxiliary problems have a bigger number of degrees of freedom than the scalar, the performance of reductions from general position to integrable nonlinear equations is more difficult; all this leads to the need of consideration gauge transformations under some restrictions, manifestly the gauge-invariant description of integrable nonlinear equations in this case is far from completion and requires additional research work.

\section*{Acknowledgment}

This research was supported by Novosibirsk State Technical University scientific grant of fundamental researches in 2007--2008 years.

\end{document}